\documentclass[10pt,journal,compsoc]{IEEEtran}

\usepackage{array}
\usepackage{mdwtab}
\usepackage{rotating}
\usepackage{hyperref}
\usepackage{xcolor}

\ifCLASSOPTIONcompsoc

  \usepackage[nocompress]{cite}
\else
  \usepackage{cite}
\fi

\begin{document}

\title{EEG-based Brain-Computer Interfaces (BCIs): A Survey of Recent Studies on Signal Sensing Technologies and Computational Intelligence Approaches and Their Applications}

\author{Xiaotong Gu,~\IEEEmembership{}
        Zehong Cao*,~\IEEEmembership{Member,~IEEE,}
        Alireza Jolfaei,~\IEEEmembership{Member,~IEEE,}
        Peng Xu,~\IEEEmembership{Member,~IEEE, }
        Dongrui Wu,~\IEEEmembership{Senior Member,~IEEE, }%
        Tzyy-Ping Jung,~\IEEEmembership{Fellow,~IEEE, }%
        and~Chin-Teng Lin,~\IEEEmembership{Fellow,~IEEE}%
        
\IEEEcompsocitemizethanks{\IEEEcompsocthanksitem X. Gu and Z. Cao are with the University of Tasmania, Australia.
\IEEEcompsocthanksitem A. Jolfaei is with the Macquarie University, Australia.
\IEEEcompsocthanksitem P. Xu is with the University of Electronic Science and Technology of China, China.
\IEEEcompsocthanksitem D. Wu is with the Huazhong University of Science and Technology, China.
\IEEEcompsocthanksitem T.P Jung is with the University of California, San Diego, USA.
\IEEEcompsocthanksitem C.T. Lin is with the University of Technology Sydney, Australia.

$*$ Corresponding to Email: Zehong.cao@utas.edu.au}% <-this % stops an unwanted space

\thanks{}}

\markboth{}%
{Shell \MakeLowercase{\textit{et al.}}: Bare Demo of IEEEtran.cls for Computer Society Journals}

\IEEEtitleabstractindextext{%
\begin{abstract}
Brain-Computer Interface (BCI) is a powerful communication tool between users and systems, which enhances the capability of the human brain in communicating and interacting with the environment directly. Advances in neuroscience and computer science in the past decades have led to exciting developments in BCI, thereby making BCI a top interdisciplinary research area in computational neuroscience and intelligence. Recent technological advances such as wearable sensing devices, real-time data streaming, machine learning, and deep learning approaches have increased interest in electroencephalographic (EEG) based BCI for translational and healthcare applications. Many people benefit from EEG-based BCIs, which facilitate continuous monitoring of fluctuations in cognitive states under monotonous tasks in the workplace or at home. In this study, we survey the recent literature of EEG signal sensing technologies and computational intelligence approaches in BCI applications, compensated for the gaps in the systematic summary of the past five years (2015-2019). In specific, we first review the current status of BCI and its significant obstacles. Then, we present advanced signal sensing and enhancement technologies to collect and clean EEG signals, respectively. Furthermore, we demonstrate state-of-art computational intelligence techniques, including interpretable fuzzy models, transfer learning, deep learning, and combinations, to monitor, maintain, or track human cognitive states and operating performance in prevalent applications. Finally, we deliver a couple of innovative BCI-inspired healthcare applications and discuss some future research directions in EEG-based BCIs.
\end{abstract}
}

\maketitle

\IEEEdisplaynontitleabstractindextext

\IEEEpeerreviewmaketitle

\IEEEraisesectionheading{\section{Introduction}\label{sec:introduction}}

\subsection{An overview of brain-computer interface (BCI)}

\subsubsection{What is BCI}
The research of brain-computer interface (BCI) was first released in the 1970s, addressing an alternative transmission channel without depending on the normal peripheral nerve and muscle output paths of the brain \cite{vidal1973toward}. An early concept of BCI proposed measuring and decoding brainwave signals to control prosthetic arm and carry out a desired action \cite{guger1999prosthetic}. Then a  formal definition of the term 'BCI' is interpreted as a direct communication pathway between the human brain and an external device \cite{wolpaw2012brain}. In the past decade, human BCIs have attracted a lot of attention. 

The corresponding human BCI systems aim to translate human cognition patterns using brain activities. It uses recorded brain activity to communicate the computer for controlling external devices or environments in a manner that is compatible with the intentions of humans \cite{lotte1999electroencephalography}, such as controlling a wheelchair or robot as shown in Fig. 1. There are two primary types of BCIs. The first type is active and reactive BCI. The active BCI derives pattern from brain activity, which is directly and consciously controlled by the user, independently from external events, for controlling a device \cite{fetz1999real}. The reactive BCI extracted outputs from brain activities in reaction to external stimulation, which is indirectly modulated by the user for controlling an application. The second type is passive BCI, which explores the user's perception, awareness, and cognition without the purpose of voluntary control, for enriching a human-computer interaction (HCI) with implicit information \cite{zander2011towards}. 

\begin{figure}[!t]
\centering
\includegraphics[width=3in]{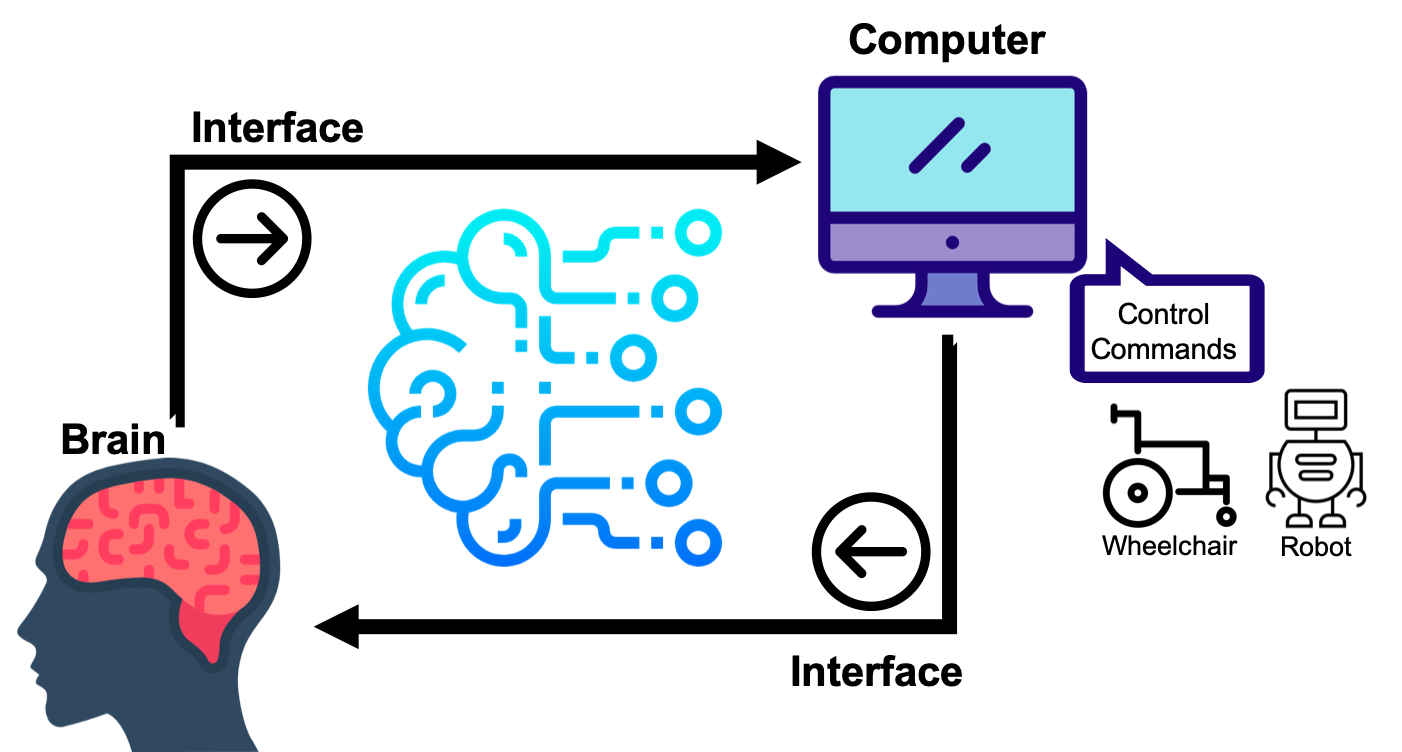}
\caption{The framework of brain-computer interface (BCI)}
\label{Fig. 1}
\end{figure}

\subsubsection{Application areas}
The promising future of BCIs has encouraged the research community to interpret brain activities to establish various research directions of BCIs. Here, we address the best-known application areas that BCIs have been widely explored and applied: (1) BCI is recognised with the potential to be an approach that uses intuitive and natural human mechanisms of processing thinking to facilitate interactions \cite{shankar2014human}. Since the common methods of traditional HCI are mostly restricted to manual interfaces and the majority of other designs are not being extensively adopted \cite{kerous2018eeg}, BCIs change how the HCI could be used in complex and demanding operational environments and could become a revolution and mainstream of HCIs for different areas such as computer-aided design (CAD) \cite{rai2016fragmentary}. Using BCIs to monitor user states for intelligent assistive systems is also substantially conducted in entertainment and health areas \cite{ienca2017intelligent}.  (2) Another area where BCI applications are broadly used is as game controllers for entertainment. Some BCI devices are inexpensive, easily portable and easy to equip, which making them feasible to be used broadly in entertainment communities. The compact and wireless BCI headsets developed for the gaming market are flexible and mobile, and require little effort to set up. Though their accuracy is not as precise as other BCI devices used in medical areas, they are still practical for game developers and successfully commercialised for the entertainment market. Some specific models \cite{mcmahan2015modality} are combined with sensors to detect more signals such as facial expressions that could upgrade the usability for entertainment applications. (3) BCIs have also been playing a significant role in neurocomputing for pattern recognition and machine learning on brain signals, and the analysis of computational expert knowledge. Recently, researches have shown \cite{badcock2015validation} \cite{finc2017transition} \cite{schmidt2018tracking} that network neuroscience approaches have been used to quantify brain network reorganisation from different varieties of human learning. The results of these studies indicate the optimisation of adaptive BCI architectures and the prospective to reveal the neural basis and future performance of BCI learning. (4) For the healthcare field, brainwave headset, which could collect expressive information with the software development kit provided by the manufacturer, has also been utilised to facilitate severely disabled people to effectively control a robot by subtle movements such as moving neck and blinking \cite{siswoyo2017application}. BCI has also been used in assisting people who lost the muscular capacity to restore communication and control over devices. A broadly investigated clinical area is to implement BCI spelling devices, one well-known application of which is a P300-based speller. Building upon the P300-based speller, \cite{fallani2019network} using a BCI2000 platform \cite{jang2014development} to develop the BCI speller has a positive result on non-experienced users to use this brain-controlled spelling tool. Overall, BCIs have contributed to various fields of research. As briefed in Fig. 2, they are involved in the entertainment of game interaction, robot control, emotion recognition, fatigue detection, sleep quality assessment, and clinical fields, such as abnormal brain disease detection and prediction including seizure, Parkinson's disease, Alzheimer's disease, Schizophrenia.

\begin{figure}[!t]
\centering
\includegraphics[width=3.1in]{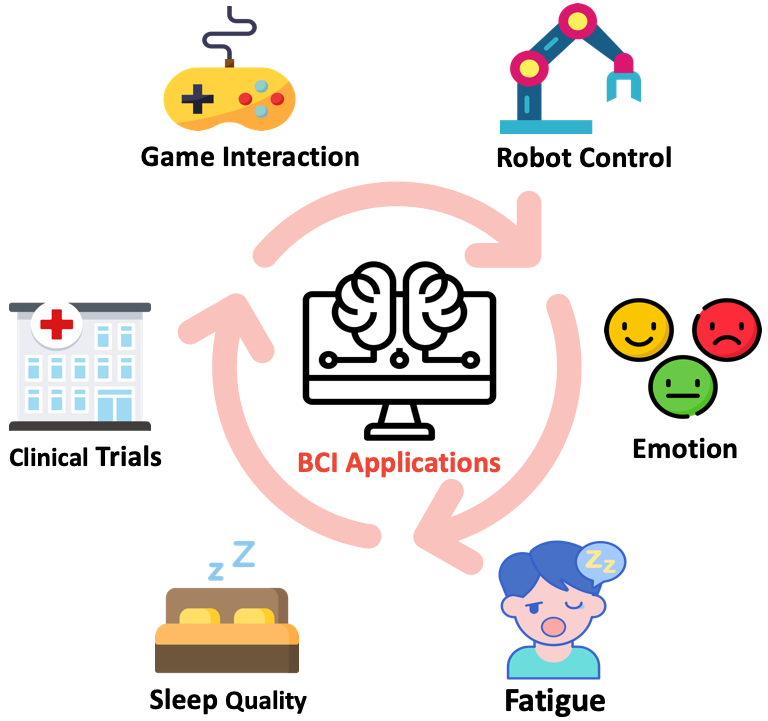}
\caption{BCI contributes to various fields of research}
\label{Fig. 2}
\end{figure}

\subsubsection{Brain imaging techniques}
Brain-sensing devices for BCI can be categorised into three groups: invasive, partially invasive, and non-invasive \cite{girouard2009distinguishing}. In terms of invasive and partially invasive devices, brain signals are collected from intracortical and electrocorticography (ECoG) electrodes with sensors tapping directly into the brain's cortex. Due to the invasive devices inserting electrodes into brain cortex, each electrode of the intracortical collection technique could provide spiking to produce the population's time developing output pattern, which causes only a slight sample of the complete set of neurons in bonded regions presented, because microelectrodes could only detect spiking when they are in the proximity of a neuron. In this case, ECoG, as an extracortical invasive electrophysiological monitoring method, uses electrodes attached under the skull. With lower surgical risk, a rather high Signal-to-Noise Ratio (SNR), and a higher spatial resolution, compared with intracortical signals of invasive devices, ECoG has a better prospect in the medical area. In specific, ECoG has a wider bandwidth to gather significant information from functional brain areas to train a high-frequency BCI system, and high SNR signals that are less prone to artefacts arising from, for instance, muscle movement and eye blink. 

Even though reliable information of cortical and neuronal dynamics could be provided by invasive or partially invasive BCIs, when considering everyday applications, the potential benefit of increased signal quality is neutralised by the surgery risks and long-term implantation of invasive devices \cite{velasco2019bci}. Recent studies started to investigate the non-invasive technology that uses external neuroimaging devices to record brain activity, including Functional Near-Infrared Spectroscopy (fNIRS), Functional Magnetic Resonance Imaging (fMRI), and Electroencephalography (EEG). To be specific, fNIRS uses near-infrared (NIR) light to assess the aggregation level of oxygenated hemoglobin and deoxygenated-hemoglobin. fNIRS depends on hemodynamic response or blood-oxygen-level-dependent (BOLD) response to formulate functional neuroimages \cite{arichi2012development}. Because of the power limits of the light and spatial resolution, fNIRS cannot be employed to measure cortical activity presented under 4cm in the brain. Also, due to the fact that blood flow changes slower than electrical or magnetic signals, Hb and deoxy-Hb have a slow and steady variation, so the temporal resolution of fNIRS is comparatively lower than electrical or magnetic signals.  fMRI monitors brain activities by assessing changes related to blood flow in brain areas, and it relies on the magnetic BOLD response, which enables fMRI to have a higher spatial resolution and collect brain information from deeper areas than fNIRS, since magnetic fields have better penetration than the NIR light. However, similar to fNIRS, the drawback of fMRI with low temporal resolutions is obvious because of the blood flow speed constraint. With the merits of relying on the magnetic response, fMRI technique also has another flaw since the magnetic fields are more prone to be distorted by deoxy-Hb than Hb molecule. The most significant disadvantage for fMRI being used in different scenarios is that it requires an expensive and heavy scanner to generate magnetic fields and the scanner is not portable and requires a lot of effort to be moved. Considering the relative rises in signal quality, reliability and mobility compared with other imaging approaches, non-invasive EEG-based devices have been used as the most popular modality for real-world BCIs and clinical use \cite{schalk2004bci2000}.

EEG signals, featuring direct measures of cortical electrical activity and high temporal resolutions, have been pursued extensively by many recent BCI studies \cite{ramadan2017brain} \cite{zander2014towards} \cite{arico2018passive}. As the most generally used non-invasive technique, EEG electrodes could be installed in a headset that is more accessible and portable for diverse occasions. EEG headsets can collect signals in several non-overlapping frequency bands (e.g. Delta, Theta, Alpha, Beta, and Gamma). This is based on the powerful intra-band connection with distinct behavioural states \cite{zhang2017deepkey}, and the different frequency bands can present diverse corresponding characteristics and patterns. Furthermore, the temporal resolution is exceedingly high on milliseconds level, and the risk on subjects is very low, compared to invasive and other non-invasive techniques that require high-intensive magnetic field exposure.  In this survey, we discussed different high portability and comparatively low-price EEG devices. A drawback of the EEG technique is that because of the limited number of electrodes, the signals have a low spatial resolution, but the temporal resolution is considerably high. When using EEG signals for BCI systems, the inferior SNR needs to be considered because objective factors such as environmental noises, and subjective factors such as fatigue status might contaminate the EEG signals. The recent research conducted to cope with this disadvantage of EEG technology is discussed in our survey as well. 

By recording small potential changes in EEG signals immediately after visual or audial stimuli appear, it is possible to observe a specific brain's response to specific stimuli events. This phenomenon is formally called Event-Related Potentials (ERPs), defined as slight voltages originated in the brain as responses to specific stimuli or events \cite{sur2009event}, which are separated into Visual Evoked Potential (VEP) and Auditory Evoked Potential (AEP). For EEG-based BCI studies, P300 wave is a representative potential response of ERP elicited in the brain cortex of a monitored subject, presenting as a positive deflection in voltage with a latency of roughly 250 to 500 ms. \cite{serby2005improved}. In specific to VEP tasks, Rapid Serial Visual Presentation (RSVP), the process of continuously presented multiple images per second at high displaying rates, is considered to have potential in enhancing human-machine symbiosis \cite{lees2018review}, and Steady-State visual evoked potentials (SSVEP) is a resonance phenomenon originating mainly in the visual cortex when a person is focusing the visual attention on a light source flickering with a frequency above 4 Hz \cite{nakanishi2017enhancing}. In addition, the psychomotor vigilance task (PVT) is a sustained-attention, reaction-timed task, measuring the speed with which subjects respond to a visual stimulus, correlates with the assessment of alertness, fatigue, or psychomotor skills \cite{li2017psychomotor}.

\subsection{Our Contributions}
Recent (2015-2019) EEG survey articles more focused on separately summarising statistical features or patterns, collecting classification algorithms, or introducing deep learning models. For example, a recent survey \cite{lotte2018review} provided a comprehensive outline regarding the latest classification algorithms utilised in EEG-based BCIs, which comprised adaptive classifiers, transfer and deep learning, matrix and tensor classifiers, and several other miscellaneous classifiers. Although \cite{lotte2018review} believes that deep learning methods have not demonstrated convincing enhancement over some state-of-the-art BCI methods, the results reviewed recently in \cite{craik2019deep} illustrated that some deep learning methods, for instance, convolutional neural networks (CNN), generative adversarial network (GAN), and deep belief networks (DBN) have outstanding performance in classification accuracy. The later review \cite{craik2019deep} synthesised performance results and the several general task groups in using deep learning for EEG classification, including sleep classifying, motor imagery, emotion recognition, seizure detection, mental workload, and event-related potential detection. In general, this review demonstrated that several deep learning methods outperformed other neural networks. However, there is no comparison between deep learning neural networks with traditional machine learning methods to prove the improvement of modern neural network algorithms in EEG-based BCIs. Another recent survey \cite{roy2019deep} systematically reviewed articles that applied deep learning to EEG in diverse domains by extracting and analysing various datasets to identify the research trend(s). It provides a comprehensive statistic evaluation for articles published between 2010 to 2018, but it does not comprise information about EEG sensors or hardware devices that collect EEG signals. Additionally, an up-to-date survey article released in early 2019 \cite{zhang2019survey} comprehensively reviewed brain signal categories for BCI and deep learning techniques for BCI applications, with a discussion of applied areas for deep learning-based BCI. While this survey provides a systematic summary over relevant publications between 2015-2019, it does not investigate thoroughly about combined machine learning, from which deep learning is originated and evolved, deep transfer learning, or the interpretable fuzzy models that are used for the non-stationary and non-linear signal processing. 

In short, the recent review articles lack a comprehensive survey in recent EEG sensing technologies, signal enhancement, relevant machine learning algorithms with interpretable fuzzy models, and deep learning methods for specific BCI applications, in addition to healthcare systems. In our survey, we aim to address the above limitations and include the recently released BCI studies in 2019. The main contributions of this study could be summarised as follow: 

•    Advances in sensors and sensing technologies.

•    Characteristics of signal enhancement and online processing.

•    Recent machine learning algorithms and the interpretable fuzzy models for BCI applications.

•    Recent deep learning algorithms and combined approaches for BCI applications.

•    Evolution of healthcare systems and applications in BCIs.

\section{Advances in Sensing Technologies}

\subsection{An overview of EEG sensors/devices}

The advanced sensor technology has enabled the development of smaller and smarter wearable EEG devices for lifestyle and related medical applications. In particular, recent advances in EEG monitoring technologies pave the way for wearable, wireless EEG monitoring devices with dry sensors. In this section, we summarise the advances of EEG devices with wet or dry sensors. We also compare the commercially available EEG devices in terms of the number of channels, sampling rate, a stationary or portable device, and covers many companies that are able to cater to the specific needs of EEG users.

\subsubsection{Wet sensor technology}

For non-invasive EEG measurements, wet electrode caps are normally attached to users' scalp with gels as the interface between sensors and the scalp. The wet sensors relying on electrolytic gels provide a clean conductive path. The application of the gel interface is to decrease the skin-electrode contact interface impedance, which could be uncomfortable and inconvenient for users and can be too time-consuming and laborious for everyday use \cite{ferree2001scalp}. However, without the conductive gel, the electrode-skin impedance cannot be measured, and the quality of measured EEG signals could be compromised.

\subsubsection{Dry sensor technology}

Because of the fact that using wet electrodes for collecting EEG data requires attaching sensors over the experimenter's skin, which is not desirable in the real-world application, the development of dry sensors of EEG devices has enhanced dramatically over the past several years \cite{mullen2015real}. One of the major advantages for dry sensors, compared with wet counterparts, is that it substantially enhances system usability \cite{liao2012biosensor}, and the headset is very easy to wear and remove, which even allows skilled users to wear it by themselves in a short time. For example, Siddharth et al. \cite{8454474} designed bio-sensors to measure physiological activities to refrain from the limitations of wet-electrode EEG equipment. These bio-sensors are dry-electrode based, and the signal quality is comparable to that obtained with wet-electrode systems but without the need for skin abrasion or preparation or the use of gels. In a follow-up study \cite{8454474}, novel dry EEG sensors that could actively filter the EEG signal from ambient electrostatic noise are designed and evaluated with ultra-low-noise and high-sampling analog to digital converter module. The study compared the proposed sensors with commercially available EEG sensors (Cognionics Inc.) in a steady-state visually evoked potentials (SSVEP) BCI task, and the SSVEP-detection accuracy was comparable between two sensors, with 74.23\% averaged accuracy across all subjects. 

Further, on the trend of wearable biosensing devices, Chi et al. \cite{chi2010dry} \cite{chi2010wireless} reviewed and designed wireless devices with dry and noncontact EEG electrode sensors. Chi et al. \cite{chi2010wireless} developed a new integrated sensor controlling the sensitive input node to attain prominent input impedance, with a complete shield of the input node from the active transistor, bond-pads, to the specially built chip package. Their experiment results, using data collected from a noncontact electrode on the top of the hair demonstrate a maximum information transfer rate at 100\% accuracy, show the promising future for dry and noncontact electrodes as viable tools for EEG applications and mobile BCIs.

Augmented BCIs (ABCIs) concept is proposed in \cite{liao2012biosensor} for everyday environments, with which signals are recorded via biosensors and processed in real-time to monitor human behaviour. An ABCI comprises non-intrusive and quick-setup EEG solutions, which requires minimal or no training, to accurately collect long-term data with the benefits of comfort, stability, robustness, and longevity. In their study of a broad range of approaches to ABCIs, developing portable EEG devices using dry electrode sensors is a significant target for mobile human brain imaging, and future ABCI applications are based on biosensing technology and devices. 

\subsection{Commercialised EEG devices}

Table I lists 21 products of 17 brands with eight attributes providing a basic overview of EEG headsets. The attribute 'Wearable' shows if the monitored human subjects could wear the devices and move around without movement constrains, which partially depends on the transmission types whether the headset devices are connected to software via Wi-Fi, Bluetooth, or other wireless techniques, or tethered connections. The numbers of channels of each EEG device could be categorised into three groups: a low-resolution group with 1 to 32 numbers; a medium-resolution group with 33 to 128 channels, and a high-resolution group with more than 128 channels. Most brands offer more than one device, therefore the numbers of channels in Table I have a wide range. The low-resolution devices mainly cover the frontal and temporal locations, some of which also deploy sensors to collect EEG signals from five locations, while the medium and high-resolution groups could cover locations of scalp more comprehensively. The numbers of channels also affect the EEG signal sampling rate of each device, with low and medium-resolution groups having a general sampling rate of 500 Hz and a high-resolution group obtaining a sampling rate of higher than 1,000 Hz. Additionally, Figure 3 presents all listed commercialised EEG devices listed in Table I.

\begin{table*}
    \centering
    \scriptsize
        \caption{An Overview of EEG Devices}
    \label{tab:my_label}
    \begin{tabular}{c|c|c|c|c|c|c|c|c}\hline
         \textbf{Brand} & \textbf{Product} & \textbf{Wearable} & \textbf{Sensors type} & \textbf{Channels No.} & \textbf{Locations} & \textbf{Sampling rate} & \textbf{Transmission} & \textbf{Weight} \\ \hline 
         \href{http://neurosky.com/}{NeuroSky} & MindWave & Yes & Dry & 1 & F & 500 Hz & Bluetooth & 90g \\
         \href{https://www.emotiv.com/}{Emotiv} & EPOC(+) & Yes & Dry & 5-14 & F, C, T, P, O & 500 Hz & Bluetooth & 125g \\
         \href{https://choosemuse.com/}{Muse} & Muse 2 & Yes & Dry & 4-7 & F, T & & Bluetooth & \\	
         \href{https://openbci.com/}{OpenBCI} & EEG Electrode Cap Kit & Yes & Wet & 8- 21 & F, C, T, P, O & & Cable & \\	
         \href{https://wearablesensing.com/}{Wearable Sensing} & DSI 24; NeuSenW & Yes & Wet; Dry & 7-21 & F, C, T, P, O & 300/600 Hz & Bluetooth & 600g \\
         \href{https://www.ant-neuro.com/}{ANT Neuro} & eego mylab / eego sports & Yes & Dry & 32 - 256 & F, C, T, P, O & Up to 16 kHz & Wi-Fi & 500g \\
         \href{http://www.neuroelectrics.com/}{Neuroelectrics} & STARSTIM; ENOBIO & Yes & Dry & 8-32 & F, C, T, P, O & 125-500 Hz & Wi-Fi; USB & \\
         \href{http://www.gtec.at/}{G.tec} & g.NAUTILUS series & Yes & Dry & 8-64 & F, C, T, P, O & 500 Hz & Wireless & 140g \\
         \href{https://www.advancedbrainmonitoring.com/}{Advanced Brain Monitoring} & B-Alert & Yes & Dry & 10-24 & F, C, T, P, O & 256Hz & Bluetooth & 110g \\
         \href{https://www.cognionics.net/}{Cognionics} & Quick & Yes & Dry & 8-30; (64-128) & F, C, T, P, O & 250/500/1k/2k Hz & Bluetooth & 610g \\
         \href{https://mbraintrain.com/}{mBrainTrain} & Smarting & Yes & Wet & 24 & F, C, T, P, O & 250-500 Hz & Bluetooth & 60g \\
         \href{https://www.brainproducts.com/productdetails.php?id=63}{Brain Products} & LiveAmp & Yes & Dry & 8-64 & F, C, T, P, O & 250/500/1k Hz & Wireless & 30g \\
         \href{https://www.brainproducts.com/productdetails.php?id=63}{Brain Products} & AntiCHapmp & Yes & Dry & 32-160 & F, C, T, P, O & 10k Hz & Wireless & 1.1kg \\
         \href{https://www.biosemi.com/}{BioSemi} & ActiveTwo & No & Wet (Gel) & 280 & F, C, T, P, O & 2k/4k/8k/16k Hz & Cable & 1.1kg \\
         \href{https://www.egi.com/}{EGI} & GES 400 & No & dry & 32-256 & F, C, T, P, O & 8k Hz & Cable & \\
         \href{http://compumedicsneuroscan.com/}{Compumedics Neuroscan} & Quick-Cap + Grael 4k & No & Wet & 32-256 & F, C, T, P, O &  & Cable & \\
         \href{http://www.mitsar-medical.com/}{Mitsar} & Smart BCI EEG Headset & Yes & Wet & 24-32 & F, C, T, P, O & 2k Hz & Bluetooth & 50g \\
         \href{http://mindo.com.tw/}{Mindo} & Mindo series & Yes & Dry & 4-64 & F, C, T, P, O &  & Wireless  & \\	
    \hline
    \multicolumn{9}{l}{Abbreviation: Frontal (F), Central (C), Temporal (T), Partial (P), and Occipital (O)} 
    \end{tabular}
\end{table*}

\begin{figure}[!t]
\centering
\includegraphics[width=3.5in]{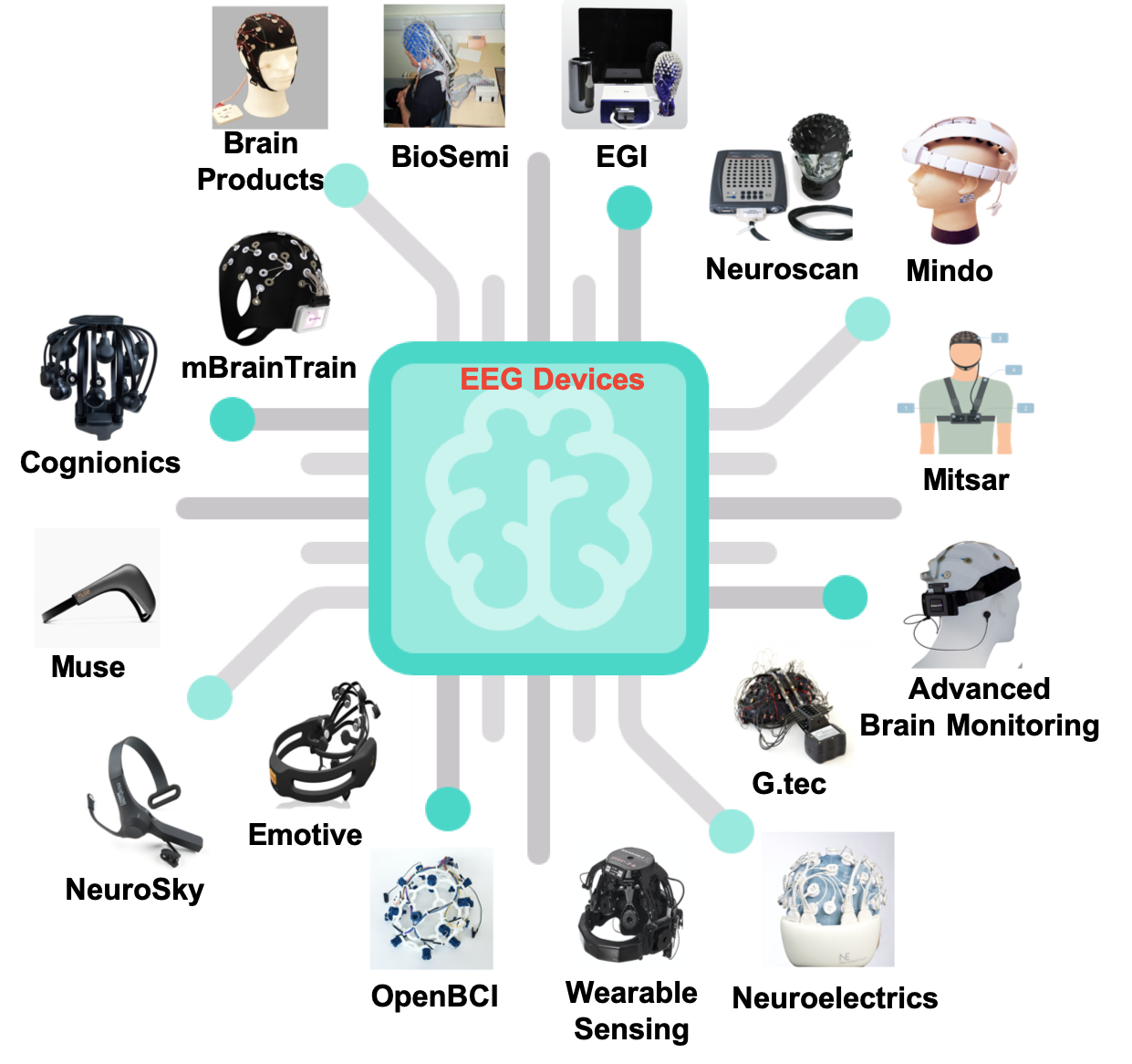}
\caption{Commercialized EEG devices for BCI applications}
\label{Fig. 3}
\end{figure}

\section{Signal Enhancement and Online Processing}

\subsection{Artefact handling}

Based on a broad category of unsupervised learning algorithms for signal enhancement, Blind Source Separation (BSS) estimates original sources and parameters of a mixing system and removes the artefact signals, such as eye blinks and movement, represented in the sources \cite{sweeney2012artefact}. There are several prevalent BSS algorithms in BCI researches, including Principal Component Analysis (PCA), Canonical Correlation Analysis (CCA) and Independent Component Analysis (ICA). PCA is one of the simplest BSS techniques, which converts correlated variables to uncorrelated variables, named principal components (PCs), by orthogonal transformation. However, the artefact components are usually correlated with EEG data and the potential of drifts are similar to EEG data, which both would cause the PCA to fail to separate the artefacts \cite{jiang2019removal}. CCA separates components from uncorrelated sources and detects a linear correlation between two multi-dimensional variables \cite{dong2015characterizing}, which has been applied in muscle artefacts removal from EEG signals. In terms of ICA, it decomposes observed signals into the independent components (ICs), and reconstruct clean signals by removing the ICs contained artefacts. ICA is the majority approach for artefacts removal in EEG signals, so we review the methods utilised ICA to support signal enhancement in the following sections.

\subsubsection{Eye blinks and movements}

Eyes blinks are more prevalent during the eyes-open condition, while rolling of the eyes might influence the eyes-closed condition. Also, the signal of eye movements located in the frontal area can affect further EEG analysis. To minimise the influence of eye contamination in EEG signals, visual inspection of artefacts and data augmentation approaches are often used to remove eye contamination. 

Artefact subspace reconstruction (ASR) is an automatic component-based mechanism as a pre-processing step, which could effectually remove large-amplitude or transient artefacts that contaminate EEG data. There are several limitations of ASR, first of which is that ASR effectually removes artefacts from the EEG signals collected from a standard 20-channel EEG device, while single-channel EEG recordings could not be applied for. Furthermore, without substantial cut-off parameters, the effectiveness of removing regularly occurring artefacts, such as eye blinks and eye movements, is limited. A possible enhancement has been proposed by using ICA-based artefact removal mechanism as a complement for ASR cleaning \cite{pion2018online}. A recent study in 2019 \cite{chang2019evaluation} also considered ICA and used an automatic IC classifier as a quantitative measure to separate brain signals and artefacts for signal enhancement. In above studies, they extended Infomax ICA \cite{lee1999independent}, and the results showed that by using an optimal ASR parameter between 20 and 30, ASR removes more eye artefacts than brain components.  

For online processing of EEG data in near real-time to remove artefacts, a combination method of online ASR, online recursive ICA, an IC classifier was proposed in \cite{pion2018online} to remove large-amplitude transients as well as to compute, categorise, and remove artefactual ICs. For the eye movement-related artefacts, the results of their proposed methods have a fluctuated saccade-related IC EyeCatch score, and the altered version of EyeCatch of their study is still not an ideal way of eliminating eye-related artefacts. 

\subsubsection{Muscle artefacts}

Contamination of EEG data by muscle activity is a well-recognized tough problem. These artefacts can be generated by any muscle contraction and stretch in proximity to the recording sites, such as the subject talks, sniffs, swallows, etc. The degree of muscle contraction and stretch will affect the amplitude and waveform of artefacts in EEG signals. In general, the common techniques that have been used to remove muscle artefacts include regression methods, Canonical Correlation Analysis (CCA), Empirical Mode Decomposition (EMD), Blind Source Separation (BSS), and EMD-BSS \cite{jiang2019removal}. A combination of the ensemble EMD (EEMD) and CCA, named EEMD-CCA, is proposed to remove muscle artefacts by \cite{chen2018novel}. By testing on real-life, semi-simulated, and simulated datasets under single-channel, few-channel and multichannel settings, the study result indicates that the EEMD-CCA method can effectively and accurately remove muscle artefacts, and it is an efficient signal processing and enhancement tool in healthcare EEG sensor networks. There are other approaches combining typical methods, such as using BSS-CCA followed by spectral-slope rejection to reduce high-frequency muscle contamination \cite{janani2018improved}, and independent vector analysis (IVA) that takes advantage of both ICA and CCA by exploiting higher-order statistics (HOS) and second-order statistics (SOS) simultaneously to achieve high performance in removing muscle artefacts \cite{chen2017independent}. A more extensive survey on muscle artefacts removal from EEG could be found in \cite{jiang2019removal}.

\subsubsection{Introducing toolbox of signal enhancement}

As one of the most widely used Matlab toolbox for EEG and other electrophysiological data processing, EEGLAB is developed by Swartz Center for Computational Neuroscience, in which provides an interactive graphic user interface for users to apply ICA, time/frequency analysis (TFA) and standard averaging methods to the recorded brain signals. EEGLAB extensions, previously called EEGLAB plugins, are the toolboxes that provide data processing and visualization functions for the EEGLAB users to process the EEG data. At the time of writing, there are 106 extensions available on the EEGLAB website, with a broad functional range including importing data, artefact removal, feature detection algorithms, etc. For example, many extensions are developed for artefact removal. Automatic artefact Removal (AAR) toolbox is for automatic EEG ocular and muscular artefact removal; Cochlear implant artefact correction (CIAC), as its name, is an ICA-based tool particularly for correcting electrical artefacts arising from cochlear implants; Multiple Artefact Rejection Algorithm (MARA) toolbox uses EEG features in the temporal, spectral and spatial domains to optimize a linear classifier to solve the component-reject vs. -accept problem and to remove loss electrodes; ‘icablinkmetrics’ toolbox aims at selecting and removing ICA components associated with eyeblink artefacts using time-domain methods. Some of the toolboxes have more than one major function, such as artefact rejection and pre-processing: “clean\_rawdata” is a suite of pre-processing methods including ASR for correcting and cleaning continuous data; ARfitStudio could be applied to quickly and intuitively correct event-related spiky artefacts as the first step of data pre-processing using “ARfit”. ADJUST identifies and removes artefact independent components automatically without affecting neural sources or data. A more comprehensive list of toolbox functions can be found on the EEGLAB website.  

\subsection{EEG Online Processing}

For neuronal information processing and BCI, the ability to monitor and analyze cortico-cortical interactions in real time is one of the trends in BCI research, along with the development of wearable BCI devices and effective approaches to remove artefacts. It is challenging to provide a reliable real-time system that could collect, extract and pre-process dynamic data with artefact rejection and rapid computation. In the model proposed by \cite{mullen2013real}, EEG data collected from a wearable, high-density (64-channel), and dry EEG device is firstly reconstructed by 3751-vertex mesh, anatomically constrained low resolution electrical tomographic analysis (LORETA), singular value decomposition (SVD) based reformulation and Automated Anatomical Labelling (AAL) before forwarded to Source Information Flow Toolbox (SIFT) and vector autoregressive (VAR) model. By applying a regularized logistic regression and testing on both simulation and real data, their proposed system is capable of real-time EEG data analysis. Later, \cite{mullen2015real} expanded their prior study by incorporating ASR for artefact removal, implementing anatomically constrained LORETA to localize sources, and adding an Alternating Direction Method of Multipliers and cognitive-state classification. The evaluation results of the proposed framework on simulated and real EEG data demonstrate the feasibility of the real-time neuronal system for cognitive state classification and identification. A subsequent study \cite{hsu2015real} aimed to present data in instantaneous incremental convergence. Online recursive least squares (RLS) whitening and optimized online recursive ICA algorithm (ORICA) are validated for separating the blind sources from high-density EEG data. The experimental results prove the proposed algorithm's capability to detect nonstationary in high-density EEG data and to extract artefact and principal brain sources quickly. Open-source Real-time EEG Source-mapping Toolbox (REST) to provide support for online artefact rejection and feature extraction are available to inspire more real-time BCI research in different domain areas. 

\section{Machine Learning and Fuzzy Models in BCI Applications}

\subsection{An overview of machine learning}

Machine learning, a subset of computational intelligence, relies on patterns and reasonings by computer systems to explore a specific task without using explicit instructions. Machine learning tasks are generally classified into several models, such as supervised learning and unsupervised learning \cite{kasabov2001evolving}. In terms of supervised learning, it is usually dividing the data into two subsets: training set (a dataset to train a model) and test set (a dataset to test the trained model) during the learning process. Supervised learning can be used for classification and regression tasks, by applying what has been learned in the training stage using labeled examples, to test the new data (testing data) for classifying types of events or predicting future events. In contrast, unsupervised machine learning is used when the data used to train is neither classified nor labelled. It contains only input data and refers to a probability function to describe a hidden structure, like grouping or clustering of data points.

Performing machine learning requires to create a model for training purpose. In EEG-based BCI applications, various types of models have been used and developed for machine learning. In the last ten years, the leading families of models used in BCIs include linear classifiers, neural networks, non-linear Bayesian classifiers, nearest neighbour classifiers and classifier combinations \cite{kotsiantis2006machine}. The linear classifiers, such as Linear Discriminant Analysis (LDA), Regularized LDA, and Support Vector Machine (SVM), classify discriminant EEG patterns using linear decision boundaries between feature vectors for each class. In terms of neural networks, they assemble layered human neurons to approximate any nonlinear decision boundaries, where the most common type in BCI applications is the Multilayer Perceptron (MLP) that typically uses only one or two hidden layers. Moving to a nonlinear Bayesian classifier, such as a Bayesian quadratic classifier and Hidden Markov Model (HMM), the probability distribution of each class is modelled, and Bayes rules are used to select the class to be assigned to the EEG patterns. Considering the physical distances of EEG patterns, the nearest neighbour classifier, such as the k nearest neighbour (kNN) algorithm, proposes to assign a class to the EEG patterns based on its nearest neighbour. Finally, classifier combinations are combining the outputs of multiple above classifiers or training them in a way that maximizes their complementarity. The classifier combinations used for BCI applications can be an enhanced, voting or stacked combination approach.

Additionally, to apply machine learning algorithms to the EEG data, we need to pre-process EEG signals and extract features from the raw data, such as frequency band power features and connectivity features between two channels \cite{daly2012brain}. Figure 4 demonstrates an EEG-based data pre-processing, pattern recognition and machine learning pipeline, to represent EEG data processing in a compact and relevant manner.

\begin{figure}[!t]
\centering
\includegraphics[width=3in]{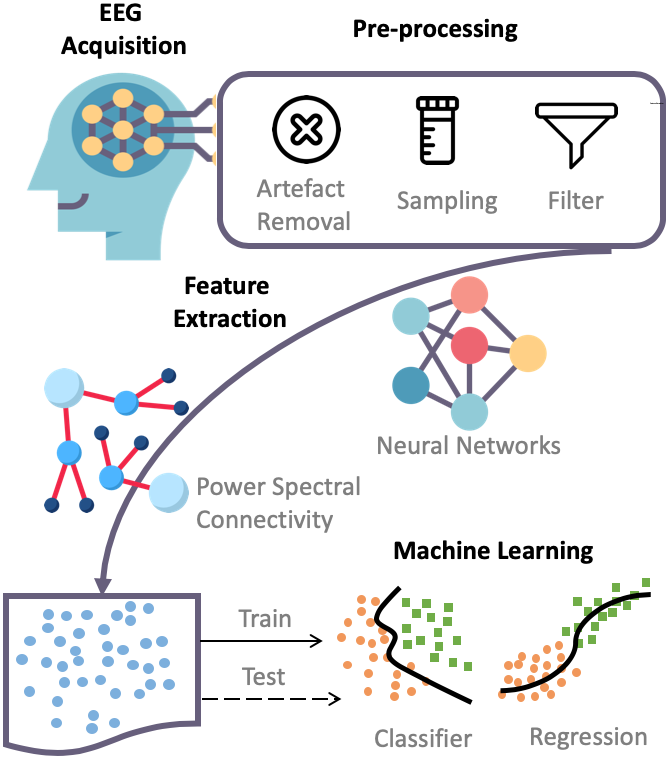}
\caption{Data pre-processing, pattern recognition and machine learning pipeline in BCIs}
\label{Fig. 4}
\end{figure}

\subsection{Transfer learning}

\subsubsection{Why we need transfer learning?}
Recently, one of the major hypotheses in traditional machine learning, such as supervised learning described above, is that training data used to train the classifier and test data used to evaluate the classifier, belong to the same feature space and follow the same probability distribution. However, this hypothesis is often violated in many applications because of human variability \cite{azab2018review}. For example, a change in EEG data distribution typically occurs when data are acquired from different subjects or across sessions and time within a subject. Also, as the EEG signals are variant and not static, extensive BCI sessions exhibit distinctive classification problems of consistency \cite{abbass2014augmented}. 

Thus, transfer learning aims at coping with data that violate this hypothesis by exploiting knowledge acquired while learning a given task for solving a different but related task. In other words, transfer learning is a set of methodologies considered for enhancing the performance of a learned classifier trained on one task (also extended to one session or subject) based on information gained while learning another one. The advances of transfer learning can relax the limitations of BCI, as it is no need to calibrate from the beginning point, less noisy for transferred information, and relying on previous usable data to increase the size of datasets.

\subsubsection{What is transfer learning?}

Transferring knowledge from the source domain to the target domain acts as bias or as a regularizer for solving the target task. Here, we provide a description of transfer learning based on the survey of Pan and Yang \cite{pan2009survey}. The source domain is known, and the target domain can be inductive (known) or transductive (unknown). The transfer learning classified under three sub-settings in accord with source and target tasks and domains, as inductive transfer learning, transductive transfer learning, and unsupervised transfer learning. 
All learning algorithms transfer knowledge to different tasks/domains, in which situation the skills should be transferred to enhance performance and avoid a negative transfer.

In the inductive transfer learning, the available labelled data of the target domain are required, while the tasks of source and target could be different from each other regardless of their domain. The inductive transfer learning could be sub-categorized into two cases based on the labelled data availability. If available, then a multitask learning method should be applied for the target and source tasks to be learned simultaneously. Otherwise, a self-taught learning technique should be deployed. In terms of transductive transfer learning, the labelled data are available in the source domain instead of the target domain, while the target and source tasks are identical regardless of the domains (of target and source). Transductive transfer learning could be sub-categorized into two cases based on whether the feature spaces between the source domain and target domain are the same. If yes, then the sample selection bias/covariance shift method should be applied. Otherwise, a domain adaptation technique should be deployed. Unsupervised transfer learning applied if available data in neither source nor target domain, while the target and source tasks are relevant but different. The target of unsupervised transfer learning is to resolve clustering, density estimation and dimensionality reduction tasks in the target domain \cite{wang2008transferred}.

\subsubsection{Where to transfer in BCIs?}
In BCIs, discriminative and stationary information could be transferred across different domains. The selection of which types of information to transfer is based on the similarity between the target and source domains. If the domains are very similar and the data sample is small, the discriminative information should be transferred; if the domains are different while there could be common information across target and source domains, stationary information should be transferred to establish more invariable systems \cite{samek2013transferring} \cite{wang2015review}. 

Domain adaption, a representative of transductive transfer learning, attempts to find a strategy for transforming the data space in which the decision rules will classify all datasets. Covariate shifting is a very related technique to domain adaptation, which is the most frequent situation encountered in BCIs. In covariate shifting, the input distributions in the training and test samples are different, while output values conditional distributions are the same \cite{shimodaira2000improving}. There exists an essential assumption - the marginal distribution of data changes from subjects (or sessions) to subjects (or sessions), and the decision rules for this marginal distribution remain unchanged. This assumption allows us to re-weight the training data from other subjects (or previous sessions) for correcting the difference in the marginal distributions in the different subjects (or sessions).

Naturally, the effectiveness of transfer learning strongly depends on how well the two circumstances are related. The transfer learning in BCI applications can be used to transfer information (a) from tasks to tasks, (b) from subjects to subjects, and (c) from sessions to sessions. As shown in Fig. 5, given a set of the training dataset (e.g., source task, subject, or session), we attempt to find a transformation space in which a training model will be beneficial to classify or predict the samples in the new dataset (e.g., target task, subject, or session).

\begin{figure}[!t]
\centering
\includegraphics[width=3.2in]{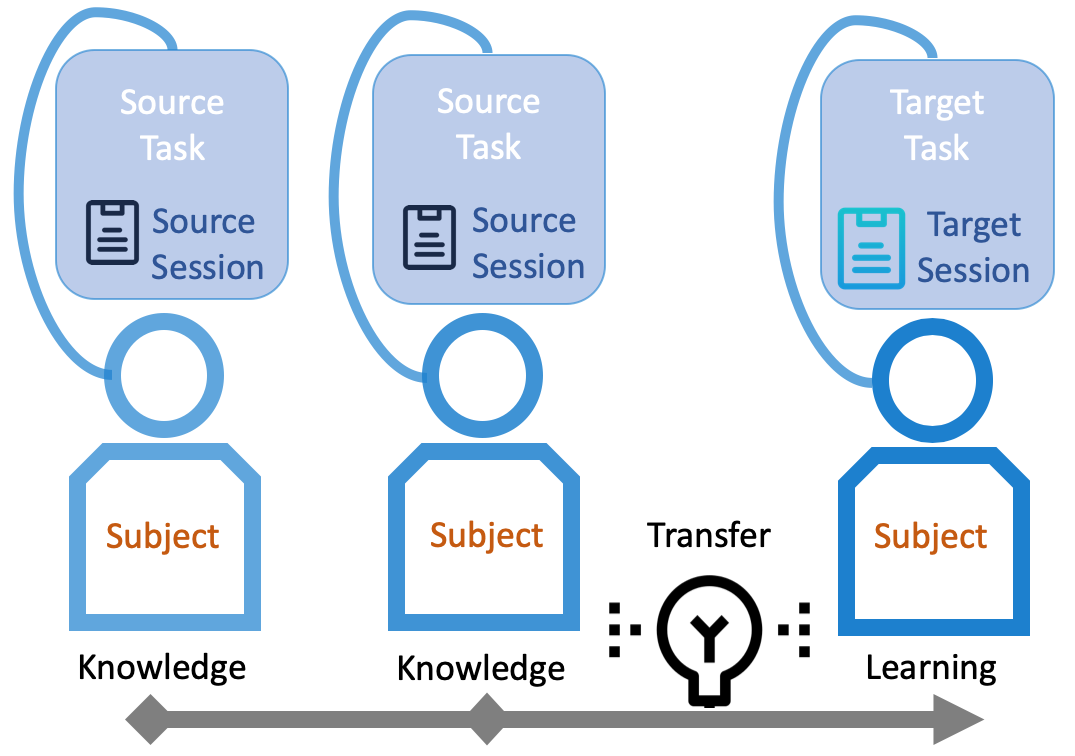}
\caption{Transfer learning in BCI}
\label{Fig. 5}
\end{figure}

\paragraph{Transfer from tasks to tasks}

In the domain of BCI where EEG signals are collected for subjects analysis, in some situations, the mental tasks and the operational tasks could be different but dependent. For instance, in a laboratory environment, a mental task is to assess the device action, such as mental subtraction and motor imagery, while the operational task is the device action itself and the performance of a device from event-related potentials. Transferring decision rules between different tasks would introduce novel signal variations and affect error-related potential that represents as a response as an error was recognized by users \cite{chavarriaga2014errare}. The study of \cite{iturrate2013task} showed that the signal variations originated from task-to-tasks transfer which substantially influenced classification feature distribution and the classifiers' performance. Other results of their study are that the accuracy based on the baseline descended when operational tasks and subtasks were generalised, while the differences of features were larger compared with non-error responses. 

\paragraph{Transfer from subjects to subjects}

For EEG-based BCIs, before applying features learned by the conventional approaches to different subjects, it requires a training period of pilot data on each new subject due to inter-subject variability \cite{chen2015high}. In the driving drowsiness detection study of Wei et al. \cite{wei2018subject}, inter- and intra-subject variability were evaluated as well as transferring models' feasibility was validated by implementing hierarchical clustering in a large-scale EEG dataset collected from many subjects. The proposed subject-to-subject transferring framework comprises a large-scale model pool, which assures sufficient data are available for positive model transferring to obtain prominent decoding performance and a small-scale baseline calibration data from the target subject as a selector of decoding models in the model pool. Without jeopardizing performance, their results in driving drowsiness detection demonstrated 90\% calibration time reduction.

In BCIs, cross-subject transfer learning could be used to decrease training data collecting time, as the least-squares transformation (LST) method proposed by Chiang et al. \cite{chiang2019cross}. The experiments conducted to validate the LST method performance of cross-subject SSVEP data showed the capability of reducing the number of training templates for an SSVEP BCI. Inter- and intra-subject transfer learning is also applied to unsupervised conditions when no labelled data is available. He and Wu \cite{he2019transfer} presented a method to align EEG trails directly in the Euclidean space across different subjects to increase the similarity. Their empirical results showed the potential of transfer learning from subjects to subjects in an unsupervised EEG-based BCI. In \cite{he2019different}, He and Wu proposed a novel different set domain adaptation approach for task-to-task and also subject-to-subject transfer, which considers a very challenging case that the source subject and the target subject have partially or completely different tasks. For example, the source subject may perform left-hand and right-hand motor imageries, whereas the target subject may perform feet and tongue motor imageries. They introduced a practical setting of different label sets for BCIs, and proposed a novel label alignment (LA) approach to align the source label space with the target label space. LA only needs as few as one labelled sample from each class of the target subject, which label alignment can be used as a preprocessing step before different feature extraction and classification algorithms, and can be integrated with other domain adaptation approaches to achieve even better performance. For applying transferring learning in BCIs, especially EEG-based BCIs, subject-to-subject transfer among the same tasks are more frequently investigated.

For subject-to-subject transfer in single-trial event-related potential (ERP) classification, Wu \cite{wu2016online} proposed both online and offline weighted adaptation regularization (wAR) algorithms to reduce the calibration effort. Experiments on a visually-evoked potential oddball task and three different EEG headsets demonstrated that both online and offline wAR algorithms are effective. Wu also proposed a source domain selection approach, which selects the most beneficial source subjects for transfer. It can reduce the computational cost of wAR by about 50\%, without sacrificing the classification performance, thus making wAR more suitable for real-time applications.

Very recently, Cui \emph{et al.} \cite{cui2019eeg} proposed a novel approach, feature weighted episodic training (FWET), to completely eliminate the calibration requirement in subject-to-subject transfer in EEG-based driver drowsiness estimation. It integrates feature weighting to learn the importance of different features, and episodic training for domain generalization. FWET does not need any labelled or unlabelled calibration data from the new subject, and hence could be very useful in plug-and-play BCIs.

\paragraph{Transfer from sessions to sessions}

The assumption of the session-to-session transfer learning in BCI is that features extracted by the training module and algorithms could be applied to a different session of a subject in the same task. It is important to evaluate what is in common among training sections for optimizing the decision distribution among different sessions.

Alamgir et al. \cite{jayaram2016transfer} reviewed several transfer learning methodologies in BCIs that explore and utilise common training data structures of several sessions to reduce training time and enhance performance. Building on the comparison and analysis of other methods in the literature, Alamgir et al. proposed a general framework for transfer learning in BCIs, which is in contrast to a general transfer learning study that focuses on domain adaptation where individual sessions feature attributes are transferred. Their framework regards decision boundaries as random variables, so the distribution of decision boundaries could be conducted from previous sessions. With an altered regression method and the consideration for feature decomposition, their experiments on amyotrophic lateral sclerosis patients using an MI BCI revealed its effectiveness in learning structure. There are also some problematic conditions of the proposed transfer learning method, including the difficulty in balancing the initialization of spatial weights, and the necessity of adding an extra loop in the algorithm for determining the spectral and spatial combination. 

In one of the latest paradigms studying imagined speech, in which a human subject imagines uttering a word without physical sound or movement, García-Salinas et al. \cite{garcia2019transfer} proposed a method to extract codewords related to the EEG signals. After a new imagined word being represented by the EEGs characteristic codewords, the new word was merged with the prior class's histograms and a classifier for transfer learning. This study implies a general trend of applying session-based transfer learning to an imagined speech domain in EEG-based BCIs.

\paragraph{Transfer from headset to headset}

Apart from the above cases of transfer learning in BCIs, ideally, a BCI system should be completely independent of any specific EEG headset, such that the user can replace or upgrade his/her headset freely, without re-calibration. This should greatly facilitate real-world applications of BCIs. However, this goal is very challenging to achieve. One step towards it is to use historical data from the same user to reduce the calibration effort on the new EEG headset.

Wu \emph{et al.} \cite{wu2016switching} proposed active weighted adaptation regularization (AwAR) for headset-to-headset transfer. It integrates wAR, which uses labelled data from the previous headset and handles class-imbalance, and active learning, which selects the most informative samples from the new headset to label. Experiments on single-trial ERP classification showed that AwAR could significantly reduce the calibration data requirement for the new headset.

\subsection{Interpretable Fuzzy Models}

Currently, machine learning methods behave like black boxes because they cannot be explained. Exploring interpretable models may be useful for understanding and improving BCI learned automatically from EEG signals, or possibly gaining new insights in BCI. Here, we collect some interpretable models from fuzzy sets and systems to estimate interpretable BCI applications.

\subsubsection{Fuzzy models for interpretability}

Zadeh suggests that all classes cannot be of clear value in the real world, so that it is very difficult to define true or false or real numbers \cite{zadeh1996fuzzy}. He introduced the concept of fuzzy sets with the definition "A fuzzy set is a set with no boundaries, and the transition from boundary to the boundary is characterized by a member function” \cite{jang1997neuro}. Using fuzzy sets allows us to provide the advantage of flexible boundary conditions, and the advances of this have applied in BCI applications, as shown in Fig. 6. 

Furthermore, a Fuzzy Inference System (FIS) also used in BCI applications to automatically extract fuzzy "If-Then" rules from the data that describe which input feature values correspond to which output category \cite{fabien2007studying}. Such fuzzy rules enable us to classify EEG patterns and interpret what the FIS has learned, as shown in Fig. 6. What is more, the fuzzy measure theory, such as fuzzy integral, as shown in Fig. 6, is suitable to apply where data fusion requires to consider possible data interactions \cite{sugeno1993fuzzy}, such as the fuzzy fusion of multiple information sources. 

Another category for interpretability is a hybrid model integrating fuzzy models to machine learning. For example, fuzzy neural networks (FNN) combine the advantages of neural networks and FIS. Its architecture is similar to the neural networks, and the input (or weight) is fuzzified \cite{buckley1994fuzzy}. The FNN recognizes the fuzzy rules and adjusts the membership function by tuning the connection weights. Especially, a Self-Organizing Neural Fuzzy Inference Network (SONFIN) is proposed by \cite{juang1999recurrent} using a dynamic FNN architecture to create a self-adaptive architecture for the identification of the fuzzy model. The advantage of designing such a hybrid structure is more explanatory because it utilises the learning capability of the neural network.

\begin{figure}[!t]
\centering
\includegraphics[width=2.2in]{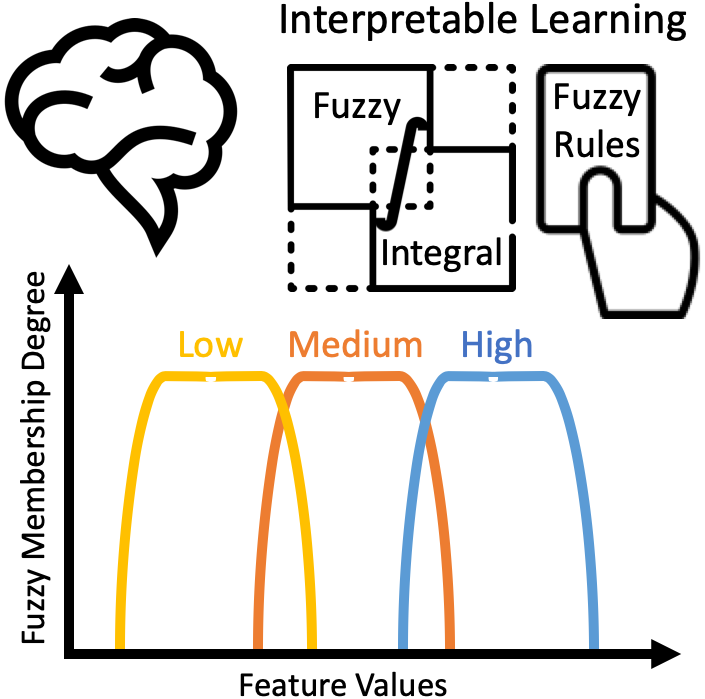}
\caption{Fuzzy sets, fuzzy rules, and fuzzy integrals for interpretability}
\label{Fig. 6}
\end{figure}

\subsubsection{EEG-based fuzzy models}

Here, we summarized up-to-date interpretable solutions based on fuzzy models for BCI systems and applications.  By using fuzzy sets, Wu et al. \cite{wu2017spatial} proposed to extend the multiclass EEG typical spatial pattern (CSP) filters from classification to regression in a large-scale sustained-attention PVT, and later further integrated them with Riemannian tangent space features for improved PVT reaction time estimation performance \cite{wu2017eeg}. Considering the advances of fuzzy membership degree, \cite{cao2017inherent} used a fuzzy membership function instead of a step function, which decreased the sensitivity of entropy values from noisy EEG signals. It did improve EEG complexity evaluation in resting state and SSVEP sessions \cite{cao2019effects}, associated with healthcare applications \cite{cao2019extraction}.

By integrating fuzzy sets with domain adaptation, \cite{wu2016driver} proposed an online weighted adaptation regularization for regression (OwARR) algorithm to reduce the amount of subject-specific calibration EEG data. Furthermore, by integrating fuzzy rules with domain adaptation, \cite{chang2017generating} generated a fuzzy rule-based brain-state-drift detector by Riemann-metric-based clustering, allowing that the distribution of the data can be observable. By adopting fuzzy integrals \cite{wu2016fuzzy}, motor-imagery-based BCI exhibited robust performance for offline single-trial classification and real-time control of a robotic arm. A follow-up work \cite{ko2019multimodal} explored the multi-model fuzzy fusion-based motor-imagery-based BCI, which also considered the possible links between EEG patterns after employing the classification of traditional BCIs. Additionally, the fusion of multiple information sources is inspired by fuzzy integrals as well, such as fusing eye movement and EEG signals to enhance emotion recognition \cite{lu2015combining}.

Moving to FNN, due to the non-linear and non-stationary characteristics of EEG signals, the application of neural networks and fuzzy logic unveils a safe, accurate and reliable detection and pattern identification. For example, a fuzzy neural detector proposed in backpropagation with a fuzzy C-means algorithm \cite{liu2015brain} and Takagi-Sugeno fuzzy measurement \cite{tsai2016takagi},  to identify the sleep stages. Furthermore, \cite{liu2015brain} proposed a recurrent self-evolving fuzzy neural network (RSEFNN) that employs an on-line gradient descent learning rule to predict EEG-based driving fatigue.

\section{Deep Learning Algorithms with BCI Applications}

Deep learning is a specific family of machine learning algorithms in which features and the classifier are jointly learned directly from data. The term ‘deep learning’ refers to the architecture of the model, which is based on a cascade of trainable feature extractor modules and nonlinearities \cite{lecun2013deep}. Owing to such a cascade, learned features are usually related to increasing levels of concepts. The representative architectures of deep learning include Convolutional Neural Networks (CNN), Generative Adversarial Network (GAN), Recurrent Neural Networks (RNN), and broad Deep Neural networks (DNN). For BCI applications, deep learning has been applied broadly compared with machine learning technology mainly because currently most machine learning research concentrates on static data which is not the optimal method for accurately categorizing the quickly changing brain signals \cite{zhang2019survey}. In this section, we introduce the spontaneous EEG applications with CNN architectures, the utilisation of GAN in recent researches, the procedure and applications of RNN, especially Long Short-Term Memory (LSTM). We also illustrate deep transfer learning extended from deep learning algorithms and transfer learning approaches, followed by exemplification of adversarial attacks to deep learning models for system testing.

\subsection{Convolutional Neural Networks (CNN)}
A Convolutional Neural Network (simplifying ConvNet or CNN) is a feedforward neural network in which information flows uni-directionally from the input to the convolution operator to the output \cite{cecotti2010convolutional}. As shown in Fig. 7, such a convolution operator includes at least three stacked layers in CNN comprising the convolutional layer, pooling layer, and fully connected layer. The convolutional layer convolves a tensor with shape, and the pooling layer streamlines the underlying computation to reduce the dimensions of the data. The fully connected layer connects every neuron in the previous layer to a new layer, resemble a traditional multi-layer perceptron neural network.

\begin{figure}[!t]
\centering
\includegraphics[width=3.5in]{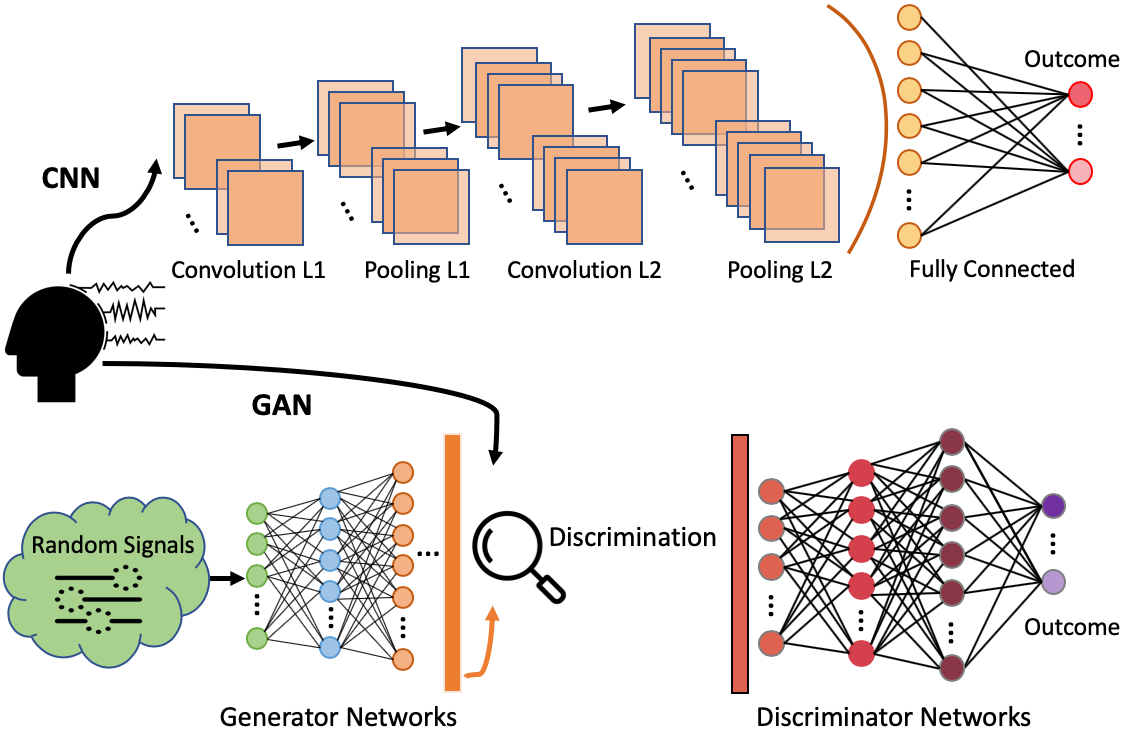}
\caption{CNN and GAN for BCI applications}
\label{Fig. 7}
\end{figure}

The nature of CNNs with stacked layers is to reduce input data to easily-identifiable formations with minimal loss, and distinctive spatial dependencies of the EEG patterns could be captured by applying CNN. For instance, CNN has been used to automatically extract signal features from epileptic intracortical data [22] and perform an automatic diagnosis to supersede the time-consuming procedure of vision examination conducted by experts [23]. In addition to this, the recent five BCI applications employing CNNs in fatigue, stress, sleep, motor imagery (MI), and emotional studies, are reviewed below.

\subsubsection{Fatigue-related EEG}

As a complex mental condition, fatigue and drowsiness are stated lack of vigilance that could lead to catastrophic incidents when the subjects are conducting activities that require high and sustained attention, such as driving vehicles. Driving fatigue detection research has been attracting considerable attention in the BCI community, \cite{chuang2018brain} \cite{cao2019multi} especially in recent years with the significant advancement of CNN in classification\cite{zeng2018eeg} \cite{cheng2018image}. An EEG-based spatial-temporal convolutional neural network (ESTCNN) was proposed by Gao et al. \cite{gao2019eeg} for driver fatigue detection. This structure was applied to eight human subjects in an experiment in which multichannel EEG signals were collected. The framework comprises a core block that extracts temporal dependencies and combines with dense layers for spatial-temporal EEG information process and classification. It is illustrated that this method could consistently decrease the data dimension in the inference process and increase reference response with computational efficiency. The results of the experiments with ESTCNN reached an accuracy rate of 97.37\% in fatigue EEG signal classification. CNN has also been used in other EEG-based fatigue recognition and evaluation applications. Yue and Wang \cite{yue2019eeg} applied various fatigue levels' EEG signals to their multi-scale CNN architecture named “MorletInceptionNet” for visual fatigue evaluation. This framework uses a space-time-frequency combined features extraction strategy to extract raw features, after which multi-scale temporal features are extracted by an inception architecture. The features are then provided to the CNN layers to classify visual fatigue evaluation. Their structure has a better performance in classification accuracy than the other five state-of-the-art methodologies, which is proof of the effectiveness of CNN in fatigue-related EEG signal processing and classification.

\subsubsection{Stress-related EEG}

Since stress is one of the leading causes of hazardous human behaviour and human errors that could cause dreadful industrial accidents, stress detection and recognition using EEG signals have become an important research area \cite{shon2018emotional}. A recent study \cite{jebelli2019mobile} proposed a new BCI framework with the CNN model and collected EEG signals from 10 construction workers whose cortisol levels, hormone-related human stress, were measured to label tasks' stress level. The result of the proposed configuration obtained the maximum accuracy rate of 86.62\%. This study proved that the BCI framework with a CNN algorithm might be the ultimate classifier for EEG-based stress recognition.

\subsubsection{Sleep-related EEG}

Sleep quality is crucial for human health in which the sleep stage classification, also called sleep scoring, has been investigated to understand, diagnose and treat sleep disorders \cite{sors2018convolutional}. Because of the lightness and portability of EEG devices, EEG is particularly suitable to recognize sleep scores. CNN has been applied to sleep stage classification by numerous studies, while the approaches of CNN using single-channel EEG are the mainstream of research investigation \cite{mousavi2019deep} \cite{rahman2018sleep} mainly due to the simplicity \cite{phan2018joint}. A single-channel EEG-based method using CNN for 5-class sleep stage conjecture in \cite{sors2018convolutional} shows competitive performance in sensible pattern detection and visualization. The significance of this research for single-channel sleep EEG processing is that it does not require feature extraction from expert knowledge or signal to learn the most suitable features to task classification end-to-end. Mousavi et al. \cite{mousavi2019deep} use a data-augmentation preprocessing method and apply raw EEG signals directly to nine convolutional layers and two fully connected layers without implicating feature extraction or feature selection. The simulation results of the study indicate the accuracy of over 93\% for the classification of 2 to 6 sleep stages classes. Furthermore, a CNN-based combined classification and prediction framework, called multitask neural networks, has also been considered for automatic sleep classifying in a recent study \cite{phan2018joint}. It is stated that this framework has the ability to generate multiple decisions, the reliability to form a final decision by aggregation, and the capability to avoid the disadvantages of the conventional many-to-one approach.

\subsubsection{MI-related EEG}
MI indicates imaging executing movement of a body part rather than conducting actual body movement in BCI systems \cite{tabar2016novel}. MI is based on the fact that brain activation will change and activate correlated brain path when actually moving a body part. The common spatial pattern (CSP) algorithm \cite{ramoser2000optimal} is an effective spatial filter that searches for a discriminative subspace to maximize one class variance and minimize the other simultaneously to classify the movement actions. CNN has also been employed to MI EEG data processing for stimulating classification performance, and there is a stream of recent research trends of combining CNN with CSP together, improving the methodology, and enhance MI classification performance \cite{korhan2019motor}. The MI classification framework proposed by Sakhavi, Guan and Yan \cite{sakhavi2018learning} presents a new data temporal representation generated from the filter-bank CSP algorithm, with CNN classification architecture. Their accuracy on the 4-class MI BCI dataset approves the usability and effectiveness of the proposed application. Olivas-Padilla and Chacon-Murguia \cite{olivas2019classification} presented two multiple MI classification methodologies that used a variation of Discriminative Filter Bank Common Spatial Pattern (DFBCSP) to extract features, after which the outcome samples have proceeded to a matrix with one or multiple pre-optimized CNN. It is stated that this proposed method could be an applicable alternative for multiple MI classification of a practical BCI application both online and offline.

\subsubsection{Emotion-related EEG}

Since it is believed that EEG contains comparatively comprehensive emotional information and better accessibility for affective research, while CNN holds the capacity to take spatial information into account with two-dimensional filters, CNN-based deep learning algorithm has been applied to EEG signals for emotion recognition and classification in numerous recent studies \cite{ko2018brief} \cite{ wang2019design} \cite{yang2018emotion} \cite{lee2018eeg}. Six basic emotional states that could be recognized and classified by using EEG signals \cite{broek2013ubiquitous} \cite{lin2010eeg}, including joy, sadness, surprise, anger, fear, and disgust, while the emotions could also be simply categorized in binary classification as positive or negative \cite{moon2018convolutional}. To apply EEG signals to CNN-based modules, EEG signals could be directly introduced to the modules, or to extract diverse entropy and power spectral density (PSD) features as the input of the models. Three connectivity features extracted from EEG signals, phase-locking value (PLV), Pearson correlation coefficient (PCC) and phase lag index (PLI), were examined in \cite{moon2018convolutional} to the proposed three different CNN structures. A popular EEG-based emotion classification database DEAP \cite{koelstra2011deap} was applied to the framework, and the PSD features performance was enhanced by the connectivity features, with PLV matrices obtaining 99.72\% accuracy utilizing CNN-5. Further, on this, dynamical graph CNN (DGCNN) has also been proposed for multichannel EEG emotion recognition. In the study of Song et al. \cite{song2018eeg}, the presented DGCNN method uses a graph to model EEG features by learning intrinsic correlations between each EEG channel to produce an adjacency matrix, which is then applied to learn more discriminative features. The experiments conducted in the SJTU emotion EEG dataset (SEED) \cite{zheng2015investigating} and the DREAMER dataset \cite{katsigiannis2017dreamer} achieved recognition accuracy rate at 90.4\% and 86.23\% respectively. 

\subsection{Generative Adversarial Networks (GAN)}

\subsubsection{GAN for data augmentation}

In classification tasks, a substantial amount of real-world data is required for training machine learning and deep learning modules, and in some cases, there are limitations of acquiring enough amount of real data or simply the investment of time and human resources could be too overwhelming. Proposed in 2014 and becoming more active in recent years, GAN is mainly used data augmentation to address the question of how to generate artificial natural looking samples to mimic real-world data via implying generative models, so that unrevealed training data sample number could be increased \cite{hartmann2018eeg}.

GAN includes two synchronously trained neural networks, “generator networks” and  “discriminator networks” as shown in Fig. 7. The “generator networks” can capture the input data distribution and aim to generate fake sample data, and the “discriminator networks” can distinguish whether the sample comes from the true training data. These two neural networks aim to generate an aggregation of samples from the pre-trained generator and to employ the samples for additional functions such as classification. 

\subsubsection{EEG data augmentation}

The significance of applying GAN for EEG is that it could address the major practical issue of insufficient training data. Abdelfattah, Abdelrahman and Wang \cite{abdelfattah2018augmenting} proposed a novel GAN model that learns statistical characteristics of the EEG and increases datasets to improve classification performance. Their study showed that the method outperforms other generative models dramatically. The Wasserstein GAN with gradient penalty (WGAN-GP) proposed by Panwar et al. \cite{panwar2019modeling} incorporates a BCI classifier into the framework to synthesize EEG data and simulate time-series EEG data. WGAN-GP was applied to event-related classification and perform task classification with the Class-Conditioned WGAN-CP. GAN has also been used in EEG data augmentation for improving recognition performance, such as emotion recognition. The framework presented in \cite{luo2018eeg} was built upon a  conventional GAN, named Conditional Wasserstein GAN (CWGAN), to enhance EEG-based emotion recognition. The high-dimensional EEG data generated by the proposed GAN framework was evaluated by three indicators to ensure high-quality synthetic data are appended to manifold supplement. The positive experiment results on SEED and DEAP emotion recognition datasets proved the effectiveness of the CWGAN model. A conditional Boundary Equilibrium GAN based EEG data augmentation method \cite{luo2019gan} for artificial differential entropy features generation was also proven to be effective in improving multimodal emotion recognition performance. 

As a branch of deep learning, GAN has been employed to generate super-resolution image copies from low-resolution images. A GAN-based deep EEG super-resolution method proposed by Corley and Huang \cite{corley2018deep} is a novel approach to generate high spatial resolution EEG data from low-resolution EEG samples via producing unsampled data from different channels. This framework could address the limitation of insufficient data collected from low-density  EEG devices by interpolating multiple missing channels effectively. 

To the best of our knowledge, in contrast with CNN, GAN was comparatively less studied in BCIs. One major reason is that the feasibility of using GAN for generating time sequence data is yet to be fully evaluated \cite{fahimi2019towards}. In the investigation of GAN performance in producing synthetic EEG signals, Fahimi et al. used real EEG signals as random input to train a CNN-based GAN to produce synthetic EEG data and then evaluate the similarities in the frequency and time domains. Their result indicates that the generated EEG data from GAN resemble the spatial, spectral, and temporal characteristics of real EEG signals. This initiates novel perspectives for future research of GAN in EEG-based BCIs.

\subsection{Recurrent Neural Networks (RNN) and Long Short-Term Memory (LSTM)}

The traditional neural networks usually are not capable of reasoning from the previous information, but RNN, inspired by human’s memory, can address this issue by adding a loop that allows information to be passed from one step of the network to the next. As shown in Fig. 8, the recurrent procedure of RNN describes a specific node $A$ in the time range $[1, t + 1]$. The node $A$ at time $t$ receives two inputs variables: $X_t$ denotes the input at time $t$ and the “backflow loop” represents the hidden state in time $[0, t – 1]$, and the node $A$ at time $t$ exports the variable $h_t$. However, the current RNN only looks at recent information to perform the present tasks in practice, so it cannot retain long-term dependencies. In this case, long short-term memory (LSTM) networks, a special kind of RNN that is capable of learning long-term dependencies, are proposed. As shown in Fig. 8, an  LSTM cell receives three inputs: the input $X$ at the current time $t$, the output $h$ of previous time $t-1$, and the “input arrows” representing the hidden state of previous time $t-1$. Then, the LSTM cell exports two outputs: the output $h$ and the hidden state (representing as the “out arrows”) of the current time $t$. The LSTM cell contains four gates, input gate, output gate, forget gate, and input modulation gate, to control the data flow by the operations and the sigmoid and tanh functions. 

Compared with traditional classification algorithms, Deep learning methods of RNN with LSTM lead to superior accuracy \cite{thomas2017deep}. Attia et al. \cite{attia2018time} presented a hybrid architecture of CNN and RNN model to categorize SSVEP signals in the time domain. Using RNN with LSTM architectures could take temporal dependence into account for EEG time-series signals and could achieve an average classification accuracy of 93.0\% \cite{hefron2017deep}. In the research of applying RNN to auditory stimulus classification, \cite{moinnereau2018classification} used a regulated RNN reservoir to classify three English vowels a, u and i. Their result showed an average accuracy rate of 83.2\% with the RNN approach outperformed a deep neural network method. A framework aiming at addressing visual object classification was proposed by \cite{spampinato2017deep} by applying RNN to EEG data invoked by visual stimulation. After discriminative brain activities are learned by the RNN, they are trained by a CNN-based regressor to project images onto the learned manifold. This automated object categorization approach achieved an accuracy rate of approximate 83\%, which proved its comparability to those empowered merely by CNN models. 

Over the past several years, the research of the RNN framework in EEG-based BCIs has increased substantially with many studies showing that the results of RNN-based methods outperform a benchmark or other traditional algorithms \cite{patnaik2017deep} or the RNN combined with other deep neural networks such as CNN to optimize performance \cite{tan2017multimodal}. The RNN framework has also been applied to other EEG-based tasks, such as identifying individuals \cite{zhang2017deepkey}, hand motion identification \cite{an2016hand}, sleep staging \cite{biswal2017sleepnet}, and emotion recognition \cite{zhang2018spatial}. It is worth noting that in \cite{biswal2017sleepnet}, the best performance model among basic machine learning, CNN, RNN, a combination of RNN and CNN (RCNN) is an RNN model with expert-defined features for sleep staging, which could inspire further research of combining the expert system with DL algorithms. Other novel framework proposals based on RNN, such as the spatial-temporal RNN (STRNN) \cite{zhang2018spatial} for feature learning integration from both temporal and spatial information of the signals, are also being explored in recent years. 

As a special kind of RNN, LSTM has also been combined with CNN algorithms for a diverse range of EEG-based tasks. For automatic sleep stage scoring, Supratak et al. \cite{supratak2017deepsleepnet} employed CNN to extract time-invariable features and an LSTM bidirectional algorithm for transitional rules learning. To predict human decisions from continuous EEG signals, \cite{hasib2018hierarchical} proposed a hierarchical LSTM model with two layers encoding local-temporal correlations and temporal correlations respectively to address non-stationarities of the EEG. 

Being able to learn sequential data and improve classification performance, LSTM could also be added to other neural networks for temporal sequential pattern detection and optimize the overall prediction accuracy for the entire framework. For temporal sleep stage classification, \cite{dong2017mixed} proposed a Mixed Neural Network with an LSTM for its capacity in learning sleep-scoring strategy automatically compared to the decision tree in which rules are defined manually.

\begin{figure}[!t]
\centering
\includegraphics[width=3.5in]{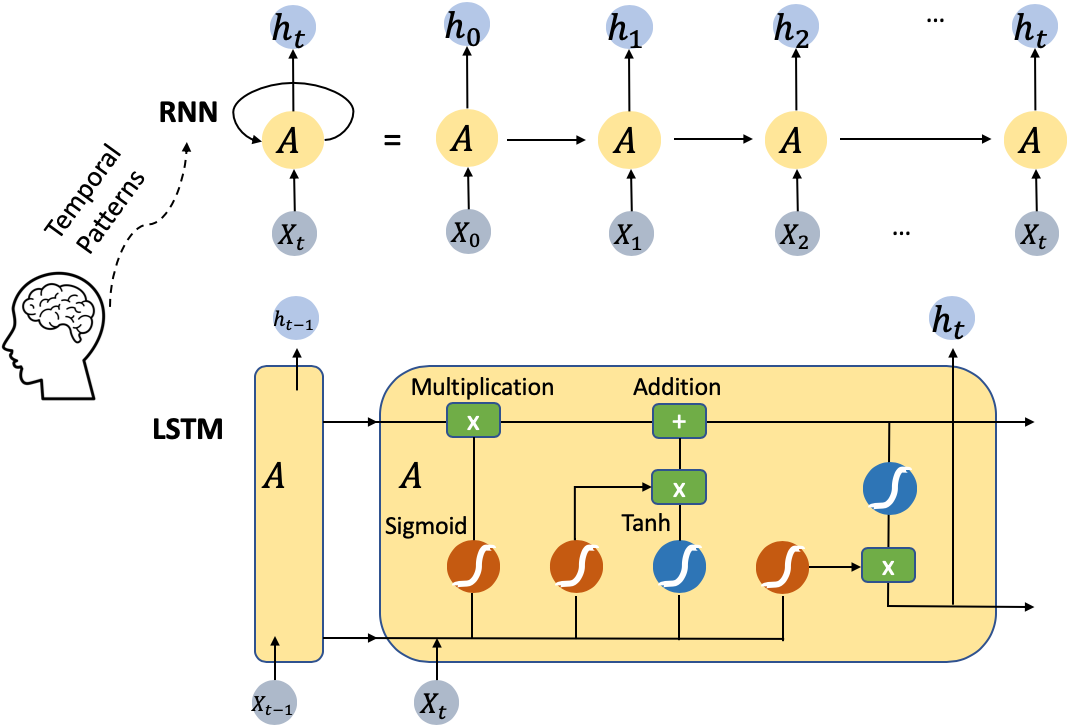}
\caption{Illustration of RNN and LSTM}
\label{Fig. 8}
\end{figure}

\subsection{Deep Transfer Learning (DTL)}
A recent survey \cite{tan2018survey} classified deep transfer learning into four categories: instance-based, mapping-based, network-based, and adversarial based deep transfer learning. In specific, the instance-based and mapping-based deep transfer learning consider instances by adjusting weights from the source domain and mapping similarities from the source to the target domains, respectively. The main benefits of applying deep learning are saving the time-consuming pre-processing while featuring engineering steps, and capturing high-level symbolic features and imperceptible dependencies simultaneously \cite{mahmud2018applications}. The realization of these two advantages is because deep learning operates straightly on raw brain signals to learn identifiable information through back-propagation and deep structures, while transfer learning is commonly applied to improve the capacity in generalization for machine learning. 

Network-based deep transfer learning reuses the parts of the network pre-trained in the source domain, such as extracting the front-layers trained by CNN. Adversarial-based deep transfer learning uses adversarial technology, such as GAN, to find transferable features that are suitable for two domains.

The combination of transfer learning and CNN has been widely used in medical applications \cite{prajapati2017classification} \cite{wimmer2016cnn} \cite{khatami2018sequential} and the general applicational purposes for instance image classification \cite{han2018new} and object recognition \cite{alexandre20163d}. In this section, we focus on transfer learning using deep neural network and its EEG-based BCI applications.

MI EEG signal classification is one of the major areas where deep transfer learning is applied. Sakhavi and Guan \cite{sakhavi2017convolutional} used a CNN model to transfer knowledge from subjects to subjects to decrease calibration time for recording data and training the model. The EEG data pipeline of deep CNN, transferring model parameters and fine-tuning on new data and using labels to regularize fine-tuning/training process, is a novel method for subjects to subjects and session to session deep transfer learning. Xu et al. \cite{xu2019deep} proposed a deep transfer CNN framework comprising a pre-trained VGG-16 CNN model and a target CNN model, between which the parameters are directly transferred, frozen and fine-tuned in the target model and MI dataset. The performance of their framework in terms of efficiency and accuracy exceed many traditional methods such as standard CNN and SVM. Dose et al. \cite{dose2018end} applied a Deep Learning approach to an EEG-based MI BCI system in healthcare, aiming to enhance the present stroke rehabilitation process. The unified model they build includes CNN layers that learn generalized features and reduce dimension, and a conventional fully connected layer for classification. By using transfer learning in this approach for adapting global classifiers to single individuals and applying raw EEG signals to this model, the results of their study reached a mean accuracy of 86.49\%, 79.25\%, and 68.51\% for datasets with two, three and four classes, respectively. For the effectiveness of alleviating training burden with transfer learning, a recent research \cite{zhu2019separated} encoded EEG features extracted from the traditional CSP by a separated channel convolutional neural network. The encoded features were then used to train a recognition network for MI classification. The accuracy of the proposed method outperformed multiple traditional machine learning algorithms.

Generally, the purpose of proposing a DTL framework as a classification strategy is to avoid time-consuming re-training and improve accuracy compared with solitary CNN and transfer learning. A deep CNN with an inter-subject transfer learning method was applied to detect attention information from the EEG time series \cite{fahimi2018inter}. DTL has also been applied to classify EEG data in imagined vowel pronunciation \cite{cooney2019optimizing}.

As introduced in the previous section, GAN, combined with transfer learning, could be rewarding in restraining domain divergence to improve domain adaptation \cite{liu2016coupled} \cite{zhu2017unpaired}. Hu et al. \cite{hu2018duplex} proposed DupGAN, a GAN framework with one encoder, one generator and two adversarial discriminators, to attain domain transformation for classification. In other streams of deep transfer learning, RNN is also applied in many EEG-based BCI studies. For EEG classification with attention-based transfer learning, \cite{tan2019attention} proposed a framework of a cross-domain TL encoder and an attention-based TL decoder with RNN for improvement in EEG classification and brain functional areas detection under different tasks. 

\subsection{Adversarial Attacks to Deep Learning Models in BCI}

Albeit their outstanding performance, deep learning models are vulnerable to adversarial attacks, where deliberately designed small perturbations, which may be hard to be detected by human eyes or computer programs, are added to benign examples to mislead the deep learning model and cause dramatic performance degradation. This phenomena was first discovered in 2014 in computer vision \cite{szegedy2013intriguing} and soon received great attention \cite{goodfellow2015explaining} \cite{kurakin2016adversarial} \cite{athalye2017synthesizing}.

Adversarial attacks to EEG-based BCIs could also cause great damage. For example, EEG-based BCIs can be used to control wheelchairs or exoskeleton for the disabled \cite{li2018feature}, where adversarial attacks could cause malfunction. In the worst case, adversarial attacks can hurt the user by driving him/her into danger on purpose. In clinical applications of BCIs in awareness/consciousness evaluation \cite{li2018feature}, adversarial attacks could lead to serious misdiagnosis.

Zhang and Wu \cite{zhang2019vulnerability} were the first to study adversarial attacks in EEG-based BCIs. They considered three different attack scenarios: 1) White-box attacks, where the attacker has access to all information of the target model, including its architecture and parameters; 2) Black-box attacks, where the attacker can observe the target model's responses to inputs; 3) Gray-box attacks, where the attacker knows some but not all information about the target model, e.g., the training data that the target model is tuned on, instead of its architecture and parameters. They showed that three popular CNN models in EEG-based BCIs, i.e., EEGNet \cite{lawhern2018eegnet}, DeepCNN and ShallowCNN \cite{schirrmeister2017deep}, can all be effectively attacked in all three scenarios.

Recently, Jiang \emph{et al.} \cite{jiang2019active} showed that query synthesis based active learning could help reduce the number of required training EEG trials in black-box adversarial attacks to the above three CNN classifiers, and Meng \emph{et al.} \cite{meng2019white} studied white-box target attacks for EEG-Based BCI regression problems, i.e., by adding a tiny perturbation, they can change the estimated driver drowsiness level or user reaction time by at least a fixed amount. Liu \emph{et al.} \cite{liu2019universal} proposes a novel total loss minimization approach to generate universal adversarial perturbations for EEG classifiers. Their approach resolved two limitations of Zhang and Wu's approaches (the attacker needs to know the complete EEG trial in advance to compute the adversarial perturbation, and the perturbation needs to be computed specifically for each input EEG trial), and hence made adversarial attacks more practical.

\section{BCI-based Healthcare Systems}
With the enhancement in the affordability and quality of EEG headsets, EEG-based BCI researches for classifying and predicting cognitive states have increased dramatically, such as tracking operators’ inappropriate states for tasks and monitoring mental health and productivity \cite{bashivan2016mental}. EEG and other brain signals such as MEG contain substantial information related to the health and disease conditions of the human brain, for instance, extracting the “slowing down” features of EEG signals could be used to categorise neurodegenerative diseases \cite{brazete2016electroencephalogram}. 

One excessive and abnormal brain disorder is epilepsy that patients suffer from recurring unprovoked seizures, which is a cause and symptom of abrupt upsurge. Clinically, EEG signals are one of the leading indicators that could be monitored and studied for seizure brain electrical activity, while EEG-based BCI researches contribute to the prediction of epilepsy. In the medical area, EEG recordings are used for screening seizures of epilepsy patients with an automated seizure detection system. The Gated Recurrent Unit RNN framework developed by \cite{talathi2017deep} showed approximate 98\% accuracy in epileptic seizure detection. Tsiouris et al. \cite{tsiouris2018long} introduced a two-layer LSTM network to assess seizure prediction performance by exploiting a broad scope of features before classification between EEG channels. The empirical results showed that the LSTM-based methodology outperforms traditional machine learning and CNN in seizure prediction performance. A novel method of EEG based automatic seizure detection was proposed by \cite{gupta2018novel} with a multi-rate filter bank structure and statistic model to optimise signal attributes for better seizer classification accuracy. For seizure detection, one of the main confusing elements is an artefact, which appears on several EEG channels that could be misinterpreted with wave and spike discharges similar to the occurrence of seizure. To optimise channel selection and accuracy for seizure detection with minimal false alarms, \cite{shah2017optimizing} proposed a CNN-LSTM algorithm to reject artefacts and optimise the framework’s performance on seizure detection. It is believed that the implementation of BCI and real-time EEG signal processing are suitable for the standard clinical application and caring for epilepsy patients \cite{alkawadri2019brain}. 

Parkinson’s disease (PD) is a progressive degradation illness classified by brain motor function, which, as an abnormal brain disease, is usually diagnosed with EEG signals. Oh et al. \cite{oh2018deep} proposed an EEG-based deep learning approach with CNN architecture as a computer-aided diagnosis system for PD detection. The positive performance result of the proposed model demonstrates its possibility in clinical usage. A specific class of RNN framework, called Echo State Networks (ESNs), was proposed by \cite{ruffini2016eeg} to classify EEG signals collected from random eye movement sleep (REM) Behavioural Disorder (RBD) patients and healthy controls subjects, while RBD is a major risk feature for neurodegenerative diseases such as PD. ESNs possess RNN’s competence of temporal pattern classification and could expedite training, and the test set performance accuracy of the proposed ESN by \cite{ruffini2016eeg} reached 85\% as an approve of effectiveness. 

As one of the most mysterious pathology, the cause of Alzheimer’s disease (AD) is still deficiently understood, and intelligent assistive technologies are believed to have the potential in supporting dementia care \cite{ienca2017intelligent}. BCI with machine learning and deep learning models is also utilised in novel researches of classifying and detecting AD, while monitoring disease effect is increasingly significant for clinical intervention. For supporting the clinical investigation, EEG signals screening of people who are vulnerable to AD could be utilised to spot the origination of AD development. With the potentiality of classification in CNN, a deep learning model with multiple convolutional-subsampling layers is proposed in \cite{morabito2016deep} and attained an averaged 80\% accuracy in categorising sets of EEG from two different classifications of subjects, one is from mild cognitive impairment subjects, a prodromal version of AD, and the other is from same age healthy control group. Simpraga et al. \cite{simpraga2017eeg} used machine learning with multiple EEG biomarkers to enhance AD classification performance and demonstrated the effectiveness of their research in improving disease identification accuracy and supports in clinical trials. 

Comparable to using deep learning models with multiple EEG biomarkers for AD classification, machine learning techniques have also been applied to EEG biomarkers for diagnosing schizophrenia. Shim et al. \cite{shim2016machine} used sensor and source level EEG features to classify schizophrenia patients and healthy controls. The result of their research indicates that the proposed tool could be promising in supporting schizophrenia diagnosis. In \cite{chu2017individual}, a modified deep learning architecture with a voting layer was proposed for individual schizophrenia classification of EEG streams. The high classification accuracy result indicates the framework’s feasibility in categorising first-episode schizophrenia patients and healthy controls. 

As non-conventional neurorehabilitation methodology, BCI has been investigated for assisting and aiding motor impairment rehabilitation, such as for patients who suffered and survived the stroke, which is a frequent high disease and generally declines patient’s mobility afterwards \cite{soekadar2015brain}. Non-invasive BCI, for instance, EEG-based technology, supports volitional transmission of brain signals to aid hand movement. BCI has great potentials in facilitating motor impairment rehabilitation via the utilisation of assistive sensation by rewarding cortical action related to sensory-motor features \cite{remsik2016review}. Frolov et al. \cite{frolov2017post} investigated the effectiveness of rehabilitation for stroke survivors with BCI training session and the research results of the participated patients indicates that combining BCI to physical therapy could enhance the results of post-stroke motor impairment rehabilitation. Other researchers also found that using BCI for motor impairment rehabilitation for post-stroke patients could help them regain body function and improve life quality \cite{clark2019brain} \cite{cervera2018brain} \cite{irimia2017brain}.

BCIs have also been employed in other healthcare areas such as investigation of migraine, pain and depressive disorders \cite{cao2015classification} \cite{cao2016resting} \cite{cao2018exploring} \cite{lin2017forehead} \cite{cao2018identifying}. Patil et al. \cite{patil2019classification} proposed an artificial neural network with supervised classifiers {}for EEG classification to detect migraine subjects. They believed that the positive results confirm that EEG-based neural network classification framework could be used for migraine detection and as a substitution for migraine diagnosis. Cao et al. \cite{cao2017estimation} \cite{cao2019extraction} presented a multi-scale relative inherent fuzzy entropy application for SSVEPs-EEG signals of two migraine phases, the pre-ictal phase before migraine attacks and the inter-ictal phase which is the baseline. The study found that for migraine patients compared with healthy controls, there are changes in EEG complexity in a repetitive SSVEP environment. Their study proved that inherent fuzzy entropy could be used in visual stimulus environments for migraine studies and has the potential in pre-ictal migraine prediction. EEG signals have also been monitored and analysed to prove the correlation between cerebral cortex spectral patterns and chronic pain intensity \cite{camfferman2017waking}. BCI based signal processing approaches could also be used in training for phantom limb pain controls by helping patients to reorganise sensorimotor cortex with the practice of hand control \cite{yanagisawa2019using}. The potential of BCI in healthcare for the general public could attract more novel researches being conducted in the near future. 

Machine learning and deep learning neural networks have been productively applied to EEG signals for various neurological disorders screening, recognition and diagnosing, and the recent researches revealed some important findings for depression detection with BCI \cite{acharya2018automated} \cite{li2019eeg}. The EEG based CAD system with CNN architecture and transfer learning method proposed by \cite{li2019eeg} indicates that the spectral information of EEG signals is critical for depression recognition while the temporal information of EEG could significantly improve accuracy for the framework. Liao et al. cite{liao2017major} proved in their research that the 8 electrodes of EEG devices from the temporal areas could provide higher accuracies in major depression detection compared with other scalp areas, which could be efficient implications for future EEG-based BCI system for depression screening. The CNN approach proposed by \cite{acharya2018automated} for EEG-Based depression screening experiments on EEG signals of depressive and normal subjects and obtain an accuracy rate of 93.5\% and 96.0\% of left and right hemisphere EEG signals respectively. Their study also confirmed the findings of a theory that depression is linked to a hyperactive right hemisphere that could inspire more novel researches for depression detection and diagnosis.  

\section{Discussion and Conclusion}
In this review, we highlighted the recent studies in the field of EEG-based by analysing over 150 studies published between 2015 and 2019 developing signal sensing technologies and applying computational intelligence approaches to EEG data. Although the advances of the dry sensor, wearable devices, the toolbox for signal enhancement, transfer learning, deep learning, or interpretable fuzzy models have lifted the performance of EEG-based BCI systems, the real-world usability challenges remain, such as the prediction or classification capability and stability in complex BCI scenarios. 

By mapping out the trend of BCI study in the past five years, we would also like to share the tendency of future directions of BCI researches. The cost-effectiveness and availability of EEG devices are attributed to the evolution of dry sensors, which in turn stimulate more research in developing enhanced sensors. The current tendency of sensor techniques focuses on augmenting signal quality with the improvement of sensor materials and emphasising user experience when collecting signals via BCI devices with comfortable sensor attachments. Fiedler et al. presented the basis for improved EEG cap designs of dry multipin electrodes in their research of polymer-based multichannel EEG electrodes system \cite{fiedler2018contact}. Their study was focused on the correlation of EEG recording quality with applied force and resulting contact pressure. Considering the comfort of wearing an EEG device for subjects, Lin et al. \cite{lin2019augmented} developed a soft, pliable pad for an augmented wire-embedded silicon-based dry-contact sensors (WSBDSs). Their study introduced copper wires in the acicular WSBDSs to ensure scalp contact on hair-covered sites and shows good performance and feasibility for applications. Chen et al. \cite{chen2018novel} proposed flexible material-based wearable sensors for EEG and other bio-signals monitoring for the tendency of smart personal devices and e-health. A closed-loop (CL) BCI method that uses biosignal simulation instant resolution could be beneficial for healthcare therapy \cite{houston2018machine}. As an example, reinforcement learning (RL) could also support improving training model accuracy in BCI applications \cite{liu2019online}. Based on the illustration of TL and DTL, the benefits of transferring extracted features and training models among subjects or tasks are apparent, such as improving training efficiency and enhancing classification accuracy. Therefore it would be encouraging to pursue experiments with adaptive EEG-based BCI training. One of the significant challenges for EEG-based technology is artefacts removal, while despite the multiple novels approaches discussed in the previous section, integrating BCI with other technical or physiological signals, which is hybrid BCI system, would be a future focus of research for improving classification accuracy and general outcomes \cite{muller2015towards} \cite{hong2017hybrid}. The scientific community is also investigating enhanced conjunction of technology and interface for HCI that is the combination of Augmented Reality (AR) and EEG-based BCI \cite{govindarajan2018immersive}. Previous researches have been inducing one popular protocol used in exogenous BCIs, SSVEPs, with visual stimulus from AR glasses such as smart glasses used in \cite{angrisani2018wearable}, and capturing the SSVEP response by measuring EEG signals to perform tasks \cite{zerafa2016real} \cite{faller2017feasibility}. With the accessibility of AR and commercialised non-invasive BCI devices, using AR and EEG devices, augmentation becomes feasible and also effective in outcomes. Finally, recent research has shown that deep learning (and even traditional machine learning) models in EEG-based BCIs are vulnerable to adversarial attacks, and there is an urgent need to develop strategies to defend such attacks.

In this paper, we systematically survey the recent advances in advances of the dry sensor, wearable devices, signal enhancement, transfer learning, deep learning, and interpretable fuzzy models for EEG-based BCIs. The various computational intelligence approaches enable us to learn reliable brain cortex features and understand human knowledge from EEG signals. In a word, we summarise the recent EEG signal sensing and interpretable fuzzy models, followed by discussing dominant transfer and deep learning for BCI applications. Finally, we overview healthcare applications and point out the open challenges and future directions.

\appendices
\section*{Acknowledgment}

Credit authors for icons made from www.flaticon.com.

\ifCLASSOPTIONcaptionsoff
  \newpage
\fi

% references section

\bibliographystyle{IEEEtran}
% argument is your BibTeX string definitions and bibliography database(s)
\bibliography{reference}

\end{document}